%% file: paper.tex
\documentclass[11pt,letterpaper,twoside]{article}

\usepackage[affil-it]{authblk}
\usepackage{fullpage}
\usepackage{fancyhdr}
\usepackage{longtable}
\usepackage{hyperref}
\usepackage{graphicx}

\PassOptionsToPackage{dvipsnames,svgnames*,x11names*}{xcolor}

\usepackage{color}
\usepackage{fancyvrb}

\DefineVerbatimEnvironment{Highlighting}{Verbatim}{commandchars=\\\{\}}
\newenvironment{Shaded}{}{}

\newcommand{\BuiltInTok}[1]{#1}

\newcommand{\CommentTok}[1]{\textcolor[rgb]{0.38,0.63,0.69}{\textit{#1}}}

\newcommand{\ConstantTok}[1]{\textcolor[rgb]{0.53,0.00,0.00}{#1}}

\newcommand{\DataTypeTok}[1]{\textcolor[rgb]{0.56,0.13,0.00}{#1}}
\newcommand{\DecValTok}[1]{\textcolor[rgb]{0.25,0.63,0.44}{#1}}

\newcommand{\FloatTok}[1]{\textcolor[rgb]{0.25,0.63,0.44}{#1}}
\newcommand{\FunctionTok}[1]{\textcolor[rgb]{0.02,0.16,0.49}{#1}}

\newcommand{\KeywordTok}[1]{\textcolor[rgb]{0.00,0.44,0.13}{\textbf{#1}}}
\newcommand{\NormalTok}[1]{#1}
\newcommand{\OperatorTok}[1]{\textcolor[rgb]{0.40,0.40,0.40}{#1}}

\newcommand{\StringTok}[1]{\textcolor[rgb]{0.25,0.44,0.63}{#1}}

\usepackage{xcolor}

\hypersetup{
  pdftitle={Toward Modern Fortran Tooling and a Thriving Developer Community},
  colorlinks=true,
  linkcolor=Maroon,
  filecolor=Maroon,
  citecolor=Blue,
  urlcolor=Blue,
pdfcreator={LaTeX via pandoc}}

\makeatletter
\def\maxwidth{\ifdim\Gin@nat@width>\linewidth\linewidth\else\Gin@nat@width\fi}
\def\maxheight{\ifdim\Gin@nat@height>\textheight\textheight\else\Gin@nat@height\fi}
\makeatother
\setkeys{Gin}{width=\maxwidth,height=\maxheight,keepaspectratio}
\makeatletter
\def\fps@figure{htbp}
\makeatother

\providecommand{\tightlist}{%
  \setlength{\itemsep}{0pt}\setlength{\parskip}{0pt}}

\makeatletter%

\renewcommand\maketitle{%
    \newpage
    \null
    \vskip 1em
    \begin{center}
    {\Large\sf\bfseries \@title \par}
    \vspace*{6pt}
    {\normalfont\normalsize \@author}
    \end{center}}

\renewcommand\section{\@startsection {section}{1}{\z@}%
                                   {-3.5ex \@plus -1ex \@minus -.2ex}%
                                   {2.3ex \@plus.2ex}%
                                   {\sf\large\bfseries}}

\renewcommand\subsection{\@startsection{subsection}{2}{\z@}%
                                     {-3.25ex\@plus -1ex \@minus -.2ex}%
                                     {1.5ex \@plus .2ex}%
                                     {\sf\normalsize\bfseries}}

\makeatother

\newlength{\cslhangindent}
\newlength{\csllabelwidth}
\newenvironment{CSLReferences}[2] 
 {
  \setlength{\parindent}{0pt}
  \ifodd #1 \everypar{\setlength{\hangindent}{\cslhangindent}}\ignorespaces\fi
  \ifnum #2 > 0
  \setlength{\parskip}{#2\baselineskip}
  \fi
 }%
 {}
\usepackage{calc}

\begin{document}

\title{Toward Modern Fortran Tooling and a Thriving Developer Community}
\input{authors}

\maketitle

\begin{abstract}
Fortran is the oldest high-level programming language that remains in
use today and is one of the dominant languages used for
compute-intensive scientific and engineering applications. However,
Fortran has not kept up with the modern software development practices
and tooling in the internet era. As a consequence, the Fortran developer
experience has diminished. Specifically, lack of a rich general-purpose
library ecosystem, modern tools for building and packaging Fortran
libraries and applications, and online learning resources, has made it
difficult for Fortran to attract and retain new users. To address this
problem, an open source community has formed on GitHub in 2019 and began
to work on the initial set of core tools: a standard library, a build
system and package manager, and a community-curated website for Fortran.
In this paper we report on the progress to date and outline the next
steps.
\end{abstract}

\hypertarget{introduction}{%
\section{Introduction}\label{introduction}}

Fortran is the oldest programming language that is still in active use
and development today. Designed and implemented at IBM by Backus et al.
(1957), it has evolved through about a dozen iterations, initially
vendor-specific, and later as an international standard (ISO/IEC JTC
1/SC 22/WG 5 2018). Today, Fortran supports multiple programming
paradigms, with strong emphasis on easy numerical calculations over
large multi-dimensional arrays and on backward compatibility. Its
Standards Committee keeps Fortran narrowly focused on the needs of
scientific and engineering application developers. The language is
natively parallel, allowing the programmer to express parallel
algorithms---on shared and distributed memory systems alike---using the
built-in language syntax and intrinsics, instead of external libraries
like the Message Passing Interface (Gropp et al. 1996). Reid, Long, and
Steidel (2020) documented the evolution of Fortran's parallel features.
Although these features have seen relatively slow implementation by the
compiler vendors, applications that use native parallel features are
becoming more common (e.g., Garain, Balsara, and Reid 2015; Mozdzynski,
Hamrud, and Wedi 2015; Shterenlikht, Margetts, and Cebamanos 2018;
Curcic 2019; Diaz et al. 2021). Today, Fortran remains the dominant
language used on top High Performance Computing (HPC) systems, with
applications to weather and climate (Powers et al. 2017; Skamarock, Ong,
and Klemp 2021), computational chemistry (Čertík, Pask, and Vackář 2013;
Apra et al. 2020), computational fluid dynamics (Sharma and Moulitsas
2017), life science (Aguilar et al. 2018; Vandenplas et al. 2020), and
economics (Harrison and Pearson 1996; Fehr and Kindermann 2018).

Fortran advances every several years through the revisions of its
standard, Fortran 2018 being the latest. However, its tooling and
library ecosystem have not benefited from the compounding effects of
community building. We speculate that this is due to the lack of
organization and communication between otherwise isolated groups of
Fortran developers. As a consequence, the Fortran user experience has
not improved to the level of other modern languages such as Python
(Sanner et al. 1999), Go (Donovan and Kernighan 2015), or Rust (Klabnik
and Nichols 2019). Specifically, general-purpose programming tools like
algorithms, containers, and good string support, typically found in
standard libraries of other languages, have been lacking. Fortran
programmers often resorted to implementing these facilities in-house,
thus reinventing the wheel at the cost of software quality and project
budget. Further, there has been no single recommended build system for
Fortran software, and most projects have been using Autotools, CMake, or
custom build scripts. A novice Fortran programmer thus needs to spend
significant amount of time learning and maintaining the build
infrastructure instead of focusing on the application or library itself.
Lack of a dedicated and user-friendly build system and package manager
has not only hindered development, but also the distribution of the
software. Improving the Fortran tooling and ecosystem of libraries is
therefore essential for the long-term well-being of the language and its
userbase.

\begin{figure}
\centering
\includegraphics{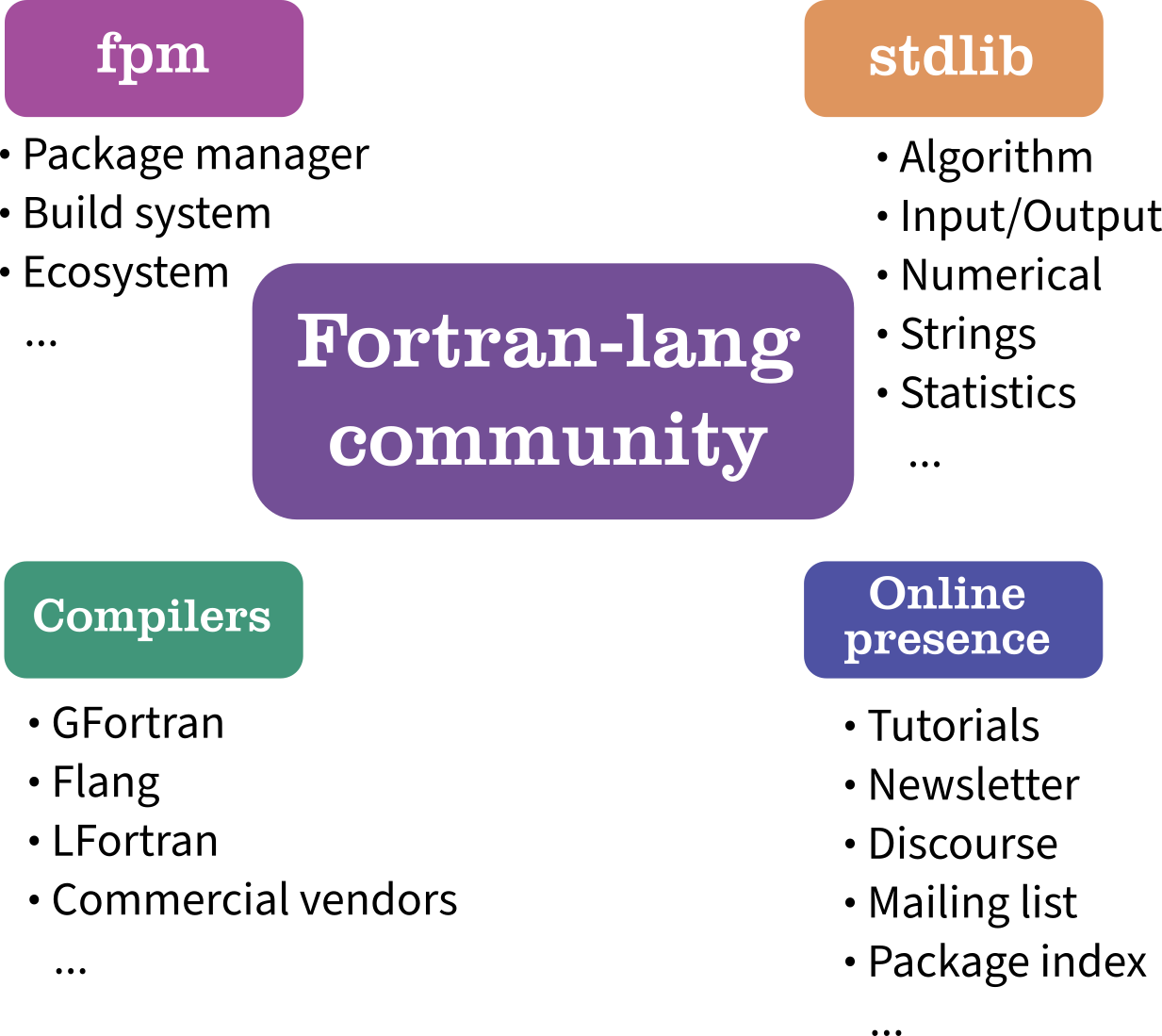}
\caption{A diagram of the Fortran-lang community and its
projects.\label{fig:community}}
\end{figure}

In this article, we describe the motivation and goals of the new open
source community, Fortran-lang, and report on the progress made so far.
We describe its three key software projects: Fortran Standard Library
(stdlib), Fortran Package Manager (fpm), and the new Fortran website and
related communication channels. A birds-eye view of each project, their
elements and themes, and how they integrate within the community is
shown on Figure \ref{fig:community}. Each of the projects we describe in
more detail in the following sections. We also describe the Fortran
proposals incubator repository, in which the Fortran-lang efforts were
first conceived. We briefly present LFortran, a modern interactive open
source compiler that will eventually integrate with other Fortran-lang
projects. Finally, we provide a brief outlook on the near-future
development for each of the projects.

\hypertarget{standard-library}{%
\section{Standard Library}\label{standard-library}}

\hypertarget{motivation}{%
\subsection{Motivation}\label{motivation}}

Unlike C or C++, The Fortran standard does not define a standard
library. Instead, it defines a number of intrinsic (built-in) modules,
procedures, and derived types (collectively called \emph{intrinsics}).
Some intrinsics are accessible through their respective intrinsic
modules (e.g.~\texttt{iso\_fortran\_env}), and others are available from
the global namespace. Fortran 2018 defines 168 intrinsic procedures, and
a smaller number of intrinsic constants and derived types. However, the
intrinsic set is limited in scope, and does not include general
algorithms, containers, and higher-level functionality commonly found in
standard libraries of languages like C++ (Stroustrup 2013), Python, and
Rust. Further, limitations in the design of strings and I/O facilities
make it awkward to carry out common tasks like loading a CSV file with
an unknown number of columns. While not impossible to do in Fortran,
such tasks are trivial to do in many other programming languages. The
lack of a standard library in Fortran thus significantly limits its ease
of use, especially for newcomers to the language.

In late 2019, it became apparent that many user-requested
functionalities could be implemented as a library, without needing to
modify the language itself. Such functionalities have been commonly
implemented ``in-house'' by developers, leading to duplication of effort
and higher risk of programming errors. Considering that a significant
fraction of Fortran application development is funded by research
grants, such duplication of effort directly incurs a significant
financial cost, and delays the completion of the application. The
traditional process of design and approval by committee, implementation
by compilers, and finally, delivery to end-users is long---often taking
more than 10 years. Instead, these features can be implemented early in
a community-curated open source library. An obvious benefit of this
approach, besides early delivery, is the opportunity to test and
discover flaws in the design of a feature before it is standardized. A
Fortran Standard Library (stdlib), designed and developed together by
the end-user community and the Standards Committee, would help mitigate
the above-mentioned problems.

\hypertarget{scope-of-stdlib}{%
\subsection{Scope of stdlib}\label{scope-of-stdlib}}

The Fortran Standard Library aims to provide a broad set of procedures,
derived types, and constants for science, engineering, mathematics, and
general-purpose programming. The science, engineering, and mathematics
scopes are similar to those of Python libraries SciPy (Virtanen et al.
2020) and NumPy (Harris et al. 2020). The general-purpose scope covers
strings, I/O, bitsets, logging, sorting, containers, and similar.

As a concrete motivation, consider the deferred-length string that was
introduced in the Fortran 2003 standard. It is a welcome addition to the
language and has a long history (e.g. L. Schonfelder (1991); J. L.
Schonfelder (1996)). While this makes life easier for Fortran
programmers, a drawback is that it is not possible to create an array of
strings where each element has its own length, independent of the other
elements, without also having to implement convenience routines.
Addressing this shortcoming of built-in Fortran strings is one of the
goals of stdlib.

\hypertarget{a-tour-of-stdlib}{%
\subsection{A tour of stdlib}\label{a-tour-of-stdlib}}

As of September 2021, stdlib implements a total of 17 modules:

\begin{itemize}
\item
  \texttt{stdlib\_ascii}: This module provides character classification
  and conversion functions for ASCII characters and strings. Programmers
  experienced with C will recognize these functions as providing the
  equivalent functionality as those in the \texttt{ctype.h} header of
  the standard C library (Plauger 1992). The classification functions
  allow to determine membership of characters in different classes:
  letters (lower-case or upper-case), digits, blank characters,
  punctuation characters, control characters (such as newline, tab, and
  carriage return), and more. Each classification function accepts a
  single character as input and returns a scalar logical value. The
  functions return \texttt{.true.} if the dummy argument \texttt{c}
  satisfies the condition described, and \texttt{.false.} if not. The
  functions available are:

  \begin{itemize}
  \tightlist
  \item
    \texttt{is\_alpha(c)} - \texttt{is\_upper(c)} or
    \texttt{is\_lower(c)} is true
  \item
    \texttt{is\_alphanum(c)} - \texttt{is\_alpha(c)} or
    \texttt{is\_digit(c)} is true
  \item
    \texttt{is\_ascii(c)} - ASCII character
  \item
    \texttt{is\_blank(c)} - space, tab
  \item
    \texttt{is\_control(c)} - control character
  \item
    \texttt{is\_digit(c)} - decimal digit
  \item
    \texttt{is\_graph(c)} - printing character except space
  \item
    \texttt{is\_hex\_digit(c)} - hexadecimal digit
  \item
    \texttt{is\_lower(c)} - lower-case letter
  \item
    \texttt{is\_octal\_digit(c)} - octal digit
  \item
    \texttt{is\_printable(c)} - printing character including space
  \item
    \texttt{is\_punctuation(c)} - printing character except space or
    letter or digit
  \item
    \texttt{is\_upper(c)} - upper-case letter
  \item
    \texttt{is\_white(c)} - space, form feed, carriage return, newline,
    tab, vertical tab
  \end{itemize}

  For the seven-bit ASCII character set, the printing characters are
  \texttt{achar(32)} to \texttt{achar(126)}; the control characters are
  \texttt{achar(0)} (NUL) to \texttt{achar(31)} (US), and
  \texttt{achar(127)} (DEL), where \texttt{achar(i)} is the Fortran
  intrinsic function that returns the character at position \texttt{i}
  in the ASCII collating sequence. For the remaining 8-bit character
  codes not in the ASCII set, the function return value is
  \texttt{.false.}. Character constants for the ASCII control characters
  are also available in the module (e.g.~\texttt{NUL}, \texttt{LF},
  \texttt{TAB}, and others). So are common character sequences such as
  lower-case and upper-case letters, digits (octal, decimal, and
  hexadecimal), as well as whitespace characters. These sequences are
  often used as the \texttt{set} argument of the intrinsic functions
  \texttt{verify} and \texttt{scan}. Finally, three functions are
  provided to convert character sequences between different cases:

  \begin{itemize}
  \tightlist
  \item
    \texttt{to\_lower(str)} - convert to lower-case
  \item
    \texttt{to\_upper(str)} - convert to upper-case
  \item
    \texttt{to\_title(str)} - capitalize first letter of each word
  \end{itemize}

  The lower- and upper-case functions are helpful to perform
  case-insensitive comparisons of character sequences and are requested
  by Fortran programmers time and again.
\item
  \texttt{stdlib\_bitsets}: This module offers two bitset types
  (\texttt{bitset\_64} and \texttt{bitset\_large}), methods for bitset
  initialization and output, unary bitset methods and binary bit
  manipulation procedures. A bitset can be seen as a memory-efficient
  representation of a rank-one array of logical values, where each
  element occupies only one bit. While the functionality of the
  \texttt{bitset\_64} type is in a way similar to that of the intrinsic
  bit manipulation procedures using integer arguments as bit sequences,
  the \texttt{bitset\_large} type is also capable of holding bitsets
  much larger than the standard integral types. Bitsets are often used
  to represent a subset of a finite set of elements. The bitwise
  \texttt{and} and bitwise \texttt{or} operators correspond to the set
  operations intersection and union, while the complement of the set can
  be found using the logical \texttt{not} operator. The following
  example program demonstrates creating and printing a bitset:

\begin{Shaded}
\begin{Highlighting}[]
\KeywordTok{program}\NormalTok{ bitset\_demo}

    \KeywordTok{use}\NormalTok{ stdlib\_bitsets, }\KeywordTok{only}\NormalTok{: bitset\_64}
    \KeywordTok{implicit} \KeywordTok{none}

    \DataTypeTok{integer}\NormalTok{, }\DataTypeTok{parameter} \DataTypeTok{::}\NormalTok{ BITSZ }\KeywordTok{=} \DecValTok{5}
    \DataTypeTok{type(bitset\_64)} \DataTypeTok{::}\NormalTok{ bitvec}
    \DataTypeTok{character(len=:)}\NormalTok{, }\DataTypeTok{allocatable} \DataTypeTok{::}\NormalTok{ bitstr}
    \DataTypeTok{integer} \DataTypeTok{::}\NormalTok{ i, cnt}

    \KeywordTok{call}\NormalTok{ bitvec\%init(BITSZ) }\CommentTok{! all zeros by default}

    \KeywordTok{call}\NormalTok{ bitvec\%set(}\DecValTok{0}\NormalTok{)}
    \KeywordTok{call}\NormalTok{ bitvec\%set(}\DecValTok{3}\NormalTok{)}
    \KeywordTok{call}\NormalTok{ bitvec\%set(}\DecValTok{4}\NormalTok{)}

    \KeywordTok{call}\NormalTok{ bitvec\%to\_string(bitstr)}
    \FunctionTok{write(*}\NormalTok{,}\StringTok{\textquotesingle{}(A)\textquotesingle{}}\FunctionTok{)}\NormalTok{ bitstr                                }\CommentTok{! prints \textquotesingle{}11001\textquotesingle{}}

    \FunctionTok{write(*}\NormalTok{,}\StringTok{\textquotesingle{}(*(I1))\textquotesingle{}}\FunctionTok{)}\NormalTok{ (bitvec\%value(i), i }\KeywordTok{=} \DecValTok{0}\NormalTok{, BITSZ}\KeywordTok{{-}}\DecValTok{1}\NormalTok{) }\CommentTok{! prints \textquotesingle{}10011\textquotesingle{}}

\KeywordTok{end program}
\end{Highlighting}
\end{Shaded}

  Bits are enumerated from the least-significant bit at \texttt{0} to
  the most significant-bit at \texttt{BITSZ\ -\ 1}. Written from left to
  right, the bitset string goes from most-significant to
  least-significant bit.
\item
  \texttt{stdlib\_error}: This module currently provides functions
  \texttt{check} and \texttt{error\_stop}.
\item
  \texttt{stdlib\_io}: This module provides high-level convenience
  wrapper functions to \texttt{open} and \texttt{close} statements. It
  also provides the functions \texttt{loadtxt} and \texttt{savetxt} that
  allow easy reading and writing of 2-dimensional arrays to and from
  text files, like the functions of the same name in the Python NumPy
  library.
\item
  \texttt{stdlib\_kinds}: This module provides a set of constants that
  define integer and real type kinds.
\item
  \texttt{stdlib\_linalg}: This module provides four functions for
  working with matrices: \texttt{diag} creates a diagonal array or
  extracts the diagonal elements from an array; \texttt{eye} returns a
  2-dimensional identity matrix of arbitrary shape;
  \texttt{outer\_product} computes the outer product of two vectors and
  returns a 2-dimensional matrix; and \texttt{trace} computes the trace
  of an input matrix.
\item
  \texttt{stdlib\_logger}: This module provides a single derived type
  \texttt{logger\_type} that allows the user to emit messages to
  standard streams or files on disk, of various kinds (debug,
  informational, warning, or error). The module also provides a global
  instance of the \texttt{logger\_type} for convenience. Here is an
  example of its use to report an error from an \texttt{allocate}
  statement:

\begin{Shaded}
\begin{Highlighting}[]
\KeywordTok{module}\NormalTok{ alloc\_mod}
    \KeywordTok{use}\NormalTok{ stdlib\_logger, }\KeywordTok{only}\NormalTok{: logger\_type}
    \KeywordTok{implicit} \KeywordTok{none}

    \DataTypeTok{type(logger\_type)} \DataTypeTok{::}\NormalTok{ logger}

    \KeywordTok{contains}

    \KeywordTok{subroutine}\NormalTok{ alloc\_check(a, }\FunctionTok{size}\NormalTok{)}
        \DataTypeTok{integer}\NormalTok{, }\DataTypeTok{allocatable}\NormalTok{, }\DataTypeTok{intent(in out)} \DataTypeTok{::}\NormalTok{ a(:)}
        \DataTypeTok{integer}\NormalTok{, }\DataTypeTok{intent(in)} \DataTypeTok{::} \FunctionTok{size}
        \DataTypeTok{integer} \DataTypeTok{::}\NormalTok{ stat}
        \DataTypeTok{character(128)} \DataTypeTok{::}\NormalTok{ errmsg, message}
        \KeywordTok{allocate}\NormalTok{(a(}\FunctionTok{size}\NormalTok{), stat}\KeywordTok{=}\NormalTok{stat, errmsg}\KeywordTok{=}\NormalTok{errmsg)}
        \KeywordTok{if}\NormalTok{ (stat }\OperatorTok{/=} \DecValTok{0}\NormalTok{) }\KeywordTok{then}
            \FunctionTok{write(}\NormalTok{ message, }\StringTok{\textquotesingle{}("Allocation of a failed with size = ", i0)\textquotesingle{}}\FunctionTok{)} \FunctionTok{size}
            \KeywordTok{call}\NormalTok{ logger \% log\_error(message,                 }\KeywordTok{\&}
                                    \KeywordTok{module=}\StringTok{\textquotesingle{}example\_mod\textquotesingle{}}\NormalTok{,    }\KeywordTok{\&}
                                    \KeywordTok{procedure=}\StringTok{\textquotesingle{}example\_sub\textquotesingle{}}\NormalTok{, }\KeywordTok{\&}
\NormalTok{                                    stat}\KeywordTok{=}\NormalTok{stat,               }\KeywordTok{\&}
\NormalTok{                                    errmsg}\KeywordTok{=}\NormalTok{errmsg)}
        \KeywordTok{end if}
    \KeywordTok{end subroutine}\NormalTok{ alloc\_check}

\KeywordTok{end module}\NormalTok{ alloc\_mod}
\end{Highlighting}
\end{Shaded}
\item
  \texttt{stdlib\_math}: This module will provide commonly used
  mathematical functions. Currently implemented are the \texttt{arange},
  \texttt{clip}, \texttt{linspace}, and \texttt{logspace} functions.
  \texttt{clip} limits an input value to a given range:

\begin{Shaded}
\begin{Highlighting}[]
\NormalTok{res }\KeywordTok{=}\NormalTok{ clip(}\FloatTok{1.22}\NormalTok{, }\FloatTok{0.}\NormalTok{, }\FloatTok{1.}\NormalTok{) }\CommentTok{! returns 1.}
\NormalTok{res }\KeywordTok{=}\NormalTok{ clip(}\KeywordTok{{-}}\FloatTok{0.5}\NormalTok{, }\FloatTok{0.}\NormalTok{, }\FloatTok{1.}\NormalTok{) }\CommentTok{! returns 0.}
\NormalTok{res }\KeywordTok{=}\NormalTok{ clip(}\FloatTok{0.75}\NormalTok{, }\FloatTok{0.}\NormalTok{, }\FloatTok{1.}\NormalTok{) }\CommentTok{! returns 0.75}
\end{Highlighting}
\end{Shaded}

  Functions \texttt{arange}, \texttt{linspace}, and \texttt{logspace}
  return sequences of evenly spaced numbers on linear or logarithmic
  scales, similar to their namesakes in NumPy.
\item
  \texttt{stdlib\_optval}: Provides a convenience function
  \texttt{optval} to handle missing optional dummy arguments. For
  example, consider a function that increments an input value by 1 by
  default, and by increment \texttt{incr} if provided:

\begin{Shaded}
\begin{Highlighting}[]
\KeywordTok{pure} \DataTypeTok{integer} \KeywordTok{function}\NormalTok{ increment(n, incr)}
    \DataTypeTok{integer}\NormalTok{, }\DataTypeTok{intent(in)} \DataTypeTok{::}\NormalTok{ n}
    \DataTypeTok{integer}\NormalTok{, }\DataTypeTok{intent(in)}\NormalTok{, }\DataTypeTok{optional} \DataTypeTok{::}\NormalTok{ incr}
    \KeywordTok{if}\NormalTok{ (}\FunctionTok{present}\NormalTok{(incr)) }\KeywordTok{then}
\NormalTok{        increment }\KeywordTok{=}\NormalTok{ n }\KeywordTok{+}\NormalTok{ incr}
    \KeywordTok{else}
\NormalTok{        increment }\KeywordTok{=}\NormalTok{ n }\KeywordTok{+} \DecValTok{1}
    \KeywordTok{end if}
\KeywordTok{end function}\NormalTok{ increment}
\end{Highlighting}
\end{Shaded}

  With \texttt{optval}, this can be written as:

\begin{Shaded}
\begin{Highlighting}[]
\KeywordTok{pure} \DataTypeTok{integer} \KeywordTok{function}\NormalTok{ increment(n, incr)}
    \KeywordTok{use}\NormalTok{ stdlib\_optval, }\KeywordTok{only}\NormalTok{: optval}
    \DataTypeTok{integer}\NormalTok{, }\DataTypeTok{intent(in)} \DataTypeTok{::}\NormalTok{ n}
    \DataTypeTok{integer}\NormalTok{, }\DataTypeTok{intent(in)}\NormalTok{, }\DataTypeTok{optional} \DataTypeTok{::}\NormalTok{ incr}
\NormalTok{    increment }\KeywordTok{=}\NormalTok{ n }\KeywordTok{+}\NormalTok{ optval(incr, }\DecValTok{1}\NormalTok{)}
\KeywordTok{end function}\NormalTok{ increment}
\end{Highlighting}
\end{Shaded}

  \texttt{optval} thus serves as syntactic sugar for easier handling of
  optional dummy arguments, until their handling is improved on the
  language level.
\item
  \texttt{stdlib\_quadrature}: This module provides different functions
  to integrate sampled values and weights for given abscissas using
  different approaches. Currently implemented are the functions to
  integrate using the Simpsons or trapezoidal rule.
\item
  \texttt{stdlib\_sorting}: This module provides efficient subroutines
  to sort arrays of intrinsic numeric and character types, as well as
  the \texttt{string\_type} from the \texttt{stdlib\_string\_type}
  module.
\item
  \texttt{stdlib\_stats}: This module provides functions to compute
  descriptive statistics from elements of integer and real arrays up to
  rank 15. Currently are implemented functions to compute Pearson
  correlations (\texttt{corr}), covariances (\texttt{cov}), means
  (\texttt{mean}), various central moments (\texttt{moment}), and
  variances (\texttt{variance}). Descriptive statistics can be computed
  along a specified dimension, and/or following a provided mask.
\item
  \texttt{stdlib\_stats\_distribution\_PRNG}: This module will provide
  pseudo-random number generators for various statistical distributions.
  Currently implemented are only functions to generate pseudo-random
  integers. Several other distributions are currently under review.
\item
  \texttt{stdlib\_string\_type}: This module contains a non-extendible
  derived type representing a character sequence, \texttt{string\_type}.
  It addresses major shortcomings of deferred length character
  variables, like the use in an \texttt{elemental} context, creating
  arrays of strings with non-uniform length, and memory safety in
  accessing the character value. Values of \texttt{string\_type} can be
  created by assignment of fixed length character values or by invoking
  the \texttt{string\_type} constructor from fixed and deferred length
  character values as well as integer and logical values.

\begin{Shaded}
\begin{Highlighting}[]
\DataTypeTok{type(string\_type)} \DataTypeTok{::}\NormalTok{ str}
\NormalTok{str }\KeywordTok{=} \StringTok{"literal"}              \CommentTok{! string contains value "literal"}
\NormalTok{str }\KeywordTok{=}\NormalTok{ string\_type()          }\CommentTok{! empty string}
\NormalTok{str }\KeywordTok{=}\NormalTok{ string\_type(}\StringTok{"literal"}\NormalTok{) }\CommentTok{! string contains value "literal"}
\NormalTok{str }\KeywordTok{=}\NormalTok{ string\_type(}\DecValTok{42}\NormalTok{)        }\CommentTok{! string contains value "42"}
\NormalTok{str }\KeywordTok{=}\NormalTok{ string\_type(}\ConstantTok{.true.}\NormalTok{)    }\CommentTok{! string contains value "T"}
\end{Highlighting}
\end{Shaded}

  All string-related intrinsic functions and operators are overloaded by
  this module to provide seamless integration with the existing use of
  deferred length character variables. Most of the overloaded functions
  are available as \texttt{elemental} to allow easy use for
  array-oriented programming. For example, scanning for the existence of
  a character set on a list of strings works naturally:

\begin{Shaded}
\begin{Highlighting}[]
\KeywordTok{program}\NormalTok{ string\_type\_demo}
    \KeywordTok{use}\NormalTok{ stdlib\_string\_type}
    \KeywordTok{implicit} \KeywordTok{none}
    \DataTypeTok{type(string\_type)}\NormalTok{, }\DataTypeTok{allocatable} \DataTypeTok{::}\NormalTok{ list(:)}

\NormalTok{    list }\KeywordTok{=} \KeywordTok{[}\NormalTok{string\_type(}\StringTok{"fortran"}\NormalTok{), string\_type(}\StringTok{"c"}\NormalTok{), string\_type(}\StringTok{"python"}\NormalTok{)}\KeywordTok{]}

    \FunctionTok{print} \KeywordTok{*}\NormalTok{, }\BuiltInTok{scan}\NormalTok{(list, }\StringTok{"ao"}\NormalTok{)  }\CommentTok{! returns [2, 0, 5]}
    \FunctionTok{print} \KeywordTok{*}\NormalTok{, }\BuiltInTok{scan}\NormalTok{(list, }\StringTok{"ao"}\NormalTok{, back}\KeywordTok{=}\ConstantTok{.true.}\NormalTok{)  }\CommentTok{! returns [6, 0, 5]}
\KeywordTok{end program}\NormalTok{ string\_type\_demo}
\end{Highlighting}
\end{Shaded}

  Furthermore, user defined derived type input and output functionality
  is provided. Repetitive tasks like reading a whole line into a
  deferred length character variable become as simple as

\begin{Shaded}
\begin{Highlighting}[]
\DataTypeTok{type(string\_type)} \DataTypeTok{::}\NormalTok{ line}
\FunctionTok{read(unit}\NormalTok{, }\FunctionTok{*)}\NormalTok{ line}
\end{Highlighting}
\end{Shaded}

  The \texttt{string\_type} is the basic building block to further
  develop string processing in stdlib.
\item
  \texttt{stdlib\_stringlist\_type}: This module provides a derived type
  that is a container for strings of varying length. Its main component
  is an allocatable array of \texttt{string\_type} instances, with
  methods to efficiently get and insert elements at arbitrary positions.
\item
  \texttt{stdlib\_strings}: This module provides procedures for string
  processing. The lack of string processing facilities has been one of
  the initial motivations behind \texttt{stdlib}.

  With numerical computations being the bread and butter of Fortran
  programmers, string processing has had a supporting role in contrast
  to other programming languages. Yet this has not stopped previous
  generations of Fortran programmers to develop useful tools including
  interpreters (e.g., MATLAB), lexical analyzers, virtual machines,
  macro processors, and even a simple programming language (Jones and
  Crabtree 1988). While repeating such ``stunts'' might be a pointless
  exercise with the powerful programming languages available today, they
  remain a testament that Fortran can be pushed further than what many
  think is possible.

  The (in)capacity of Fortran string processing facilities has been
  recognized previously by Jones and Crabtree (1988): ``Fortran is often
  maligned for its lack of facilities for character-oriented processing.
  \ldots{} The apparent deficiency of Fortran for string manipulation is
  primarily because of the methods traditionally used rather than
  because of a shortcoming of the language itself. The main shortcoming
  of Fortran for string handling is the lack of a standard library of
  routines for often-needed functions. As Fortran programmers we are
  faced with a choice: we either invest the up-front effort required to
  create our own standard library or we live with the continuing effort
  of hacking together a solution each time we are presented with similar
  problems.''

  Compared to numerical applications, which have been traditionally
  covered by vendor libraries (NAG, IMSL, HSL, MKL), string handling
  libraries are typically small, single-user driven, and dispersed.

  To address these deficiencies, the \texttt{stdlib\_string} module
  provides the following functions:

  \begin{itemize}
  \tightlist
  \item
    \texttt{chomp} - remove trailing characters, given an input set or
    substring of characters to remove;
  \item
    \texttt{count} - count the number of substring patterns in a string;
  \item
    \texttt{find} - return an n-th occurrence of a substring pattern in
    a given string;
  \item
    \texttt{padl} and \texttt{padr} - left- and right-pad a string with
    an input character;
  \item
    \texttt{reverse} - reverse a string;
  \item
    \texttt{slice} - extract characters from a defined region of the
    input string by taking strides;
  \item
    \texttt{starts\_with} and \texttt{ends\_width} - test that the
    string begins or ends with a substring;
  \item
    \texttt{strip} - remove all whitespace characters;
  \item
    \texttt{to\_string} - convert an intrinsic numeric or logical scalar
    value into a string, given an input format string (edit descriptor).
  \end{itemize}

  String facilities that remain to be implemented include subroutines
  for splitting and joining strings, pattern matching and substitution,
  tokenization and parsing, dynamic string lists, transformations
  (padding, removing spaces or other trailing characters), remaining
  character to numeric conversion routines and vice versa, among others.
\item
  \texttt{stdlib\_system}: This module will contain procedures for
  portable system programming on different platforms. Many programs need
  to get information about what is going on outside, and act upon this
  information inside of the program. Apart from the intrinsic
  \texttt{read} and \texttt{write} statements, standard Fortran provides
  limited support to execute system commands, retrieve command-line
  arguments, inspect environment variables, and interact with the file
  system. To fill this void, compiler vendors currently provide their
  own extension modules (e.g.~the Intel \texttt{ifport} module, or NAG's
  \texttt{f90\_unix\_*} modules) or offer (non-standard) built-in
  procedures to C system functions. Plans for the
  \texttt{stdlib\_system} module include procedures to retrieve
  information about the system (operating system, platform endianness)
  and user (group), list and manage files, directories, and their
  attributes, interact with terminal emulators, spawn and control
  subprocesses, and handle signals and program control, among others.

  Many of these goals can be achieved by interfacing to the respective
  system libraries using the standard C interoperability features. So
  far, the \texttt{stdlib\_system} module only offers a cross-platform
  version of the subroutine \texttt{sleep(millisec)} which causes the
  calling thread to sleep for an integer amount of milliseconds. Efforts
  are under way to provide basic file system operations, such as
  creating and removing directories, renaming files, and manipulating
  file paths.
\end{itemize}

The Fortran stdlib API documentation (https://stdlib.fortran-lang.org)
is generated by FORD (MacMackin 2018).

\hypertarget{auto-generation-of-specific-procedures}{%
\subsection{Auto-generation of specific
procedures}\label{auto-generation-of-specific-procedures}}

A significant limitation of generic programming in Fortran is that one
needs to implement many specific procedures that are to be invoked using
a generic name. Consider for example a function \texttt{mean(x)} which
returns the arithmetic average of all elements of an input array
\texttt{x}. Implementing a generic function \texttt{mean(x)} that
operates on integer or real arrays of any standard type kind (four for
integers and three for reals), and of any rank (1 through 15), requires
writing 105 specific procedures. To efficiently generate specific
procedures for any number of types, type kinds, and ranks, we use the
Python-powered preprocessor for Fortran, Fypp (Aradi 2020). For example,
the following listing shows the meta-code that defines the \texttt{mean}
function to be processed by Fypp:

\begin{verbatim}
#:for k1, t1 in RC_KINDS_TYPES
  #:for rank in RANKS
    #:set RName = rname("mean_all",rank, t1, k1)
    module function ${RName}$ (x) result(res)
      ${t1}$, intent(in) :: x${ranksuffix(rank)}$
      ${t1}$ :: res

      res = sum(x) / real(size(x, kind = int64), ${k1}$)

    end function ${RName}$
  #:endfor
#:endfor
\end{verbatim}

where the \texttt{RC\_KINDS\_TYPES} and \texttt{RANKS} variables are
defined elsewhere to specify which types, type kinds, and ranks will be
used to generate the Fortran code.

\hypertarget{supported-platforms}{%
\subsection{Supported platforms}\label{supported-platforms}}

To build the Fortran Standard Library, the following tools are required:

\begin{itemize}
\tightlist
\item
  at least a Fortran 2008 compliant compiler;
\item
  CMake version 3.14 or newer. Alternatively Make can be used;
\item
  the Python-powered preprocessor Fypp.
\end{itemize}

Using the Continuous Integration and Deployment (CI/CD) pipeline
provided by GitHub, all functions are tested at each merge of a pull
request using various compilers on various operating systems.
Specifically, stdlib is currently tested with GCC Fortran versions 9 to
11 on Ubuntu 18.04, macOS Catalina 10.15, and Windows Server 2019, and
with Intel OneAPI 2021.1 on Ubuntu 20.04 and macOS Catalina 10.15. In
addition to the compilers and OSs regularly tested by the CI/CD pipeline
of stdlib, a list of compilers and operating systems tested by stdlib's
users can be found on the GitHub README page of stdlib.

\hypertarget{fortran-package-manager}{%
\section{Fortran Package Manager}\label{fortran-package-manager}}

Among the biggest challenges that Fortran programmers deal with is
building packages, reusing them as dependencies in other packages, and
distributing them to other users. Until recently, Fortran programmers
resorted to building their packages using Autotools, CMake, hand-written
Makefiles, and even custom shell scripts. The latter approach is
especially common to large community applications that are intended to
build and run on many OSs and compiler vendors---a typical example being
the Weather Research and Forecasting model (WRF, Powers et al. (2017)).

The problem with current approaches is the lack of a well established,
recommended way to set them up so that multiple libraries work well
together. Developers resort to custom and \emph{ad hoc} workarounds to
get the build systems of different libraries to work together in their
project. They often find that it is not worth the effort, and rewrite
the build system to fit the need of their project. This leads to
duplicated effort and yet another solution that is not reusable in other
projects.

The Fortran Package Manager (fpm) is designed as a
Fortran-specific---rather than language agnostic---build system and
package manager. Its key goal is to improve the user experience by
making the development, reuse, and distribution of Fortran libraries and
applications seamless and easy. Fpm is largely modeled after Rust's
Cargo, and takes inspiration from Haskell's Stack, Node.js npm, and
others. Using and managing dependencies with these systems often require
no more than a single line in a project's configuration file. Fpm
assumes a default layout of the package to enable automatic discovery of
module and program source files. A project that follows the fpm
convention can thus be built just using the command \texttt{fpm\ build},
with minimal configuration necessary. By making the building and
packaging as easy as possible for developers, fpm's long term vision is
to lower the bar for development and distribution of packages, and thus
nurture and grow the ecosystem of modern Fortran applications and
libraries.

Fpm is written in modern Fortran, and is an fpm package itself--given a
binary executable of fpm, developers can use it to bootstrap fpm from
source. An initial prototype was written in Haskell in the Spring of
2020. Its goal was to quickly prototype and test the initial design of
the system. It implemented sufficient functionality to allow other
open-source Fortran developers to adopt fpm's packaging system, provide
feedback, and even contribute back to fpm. By the Summer of 2020 we
became confident that the initial design of fpm was likely to work, as
seen from the adoption and feedback during the prototype phase, when
Fortran developers began adopting fpm's packaging conventions in their
own projects. We began working on the production implementation of fpm
in Fortran, bringing two major boons to the project. First, a large
community of Fortran developers began contributing to fpm immediately.
Second, fpm helped motivate the development of auxiliary tools---such as
strings and file system operations---that are sorely lacking from the
Fortran ecosystem. A Fortran implementation of fpm thus helps feed back
to the Standard Library and the broader ecosystem of libraries. As more
3rd-party libraries and applications adopt the fpm convention, its value
grows thanks to the network effects---the more libraries there are to
use as dependencies, the easier it becomes to start and rapidly
prototype a new project.

\hypertarget{a-short-demo-of-fpm}{%
\subsection{A short demo of fpm}\label{a-short-demo-of-fpm}}

Using fpm should feel natural to anyone familiar with command line
operations, especially POSIX compliant tools. Executing the command
\texttt{fpm} with no arguments, or with the \texttt{-h} or
\texttt{-\/-help} options, provides a summary of its options, as
illustrated below.

\begin{verbatim}
$ fpm
Fortran Package Manager:

USAGE: fpm [ SUBCOMMAND [SUBCOMMAND_OPTIONS] ]|[--list|--help|--version]
       where SUBCOMMAND is commonly new|build|run|test

 subcommand may be one of

  build     Compile the package placing results in the "build" directory
  help      Display help
  list      Display this list of subcommand descriptions
  new       Create a new Fortran package directory with sample files
  run       Run the local package application programs
  test      Run the test programs
  update    Update and manage project dependencies
  install   Install project

 Enter "fpm --list" for a brief list of subcommand options. Enter
 "fpm --help" or "fpm SUBCOMMAND --help" for detailed descriptions.
\end{verbatim}

Using the \texttt{-h} or \texttt{-\/-help} options with a subcommand
provides additional details about the options that it supports. For
example, executing the command \texttt{fpm\ build\ -\/-help} reports the
following information:

\begin{verbatim}
$ fpm build --help
NAME
 build(1) - the fpm(1) subcommand to build a project

SYNOPSIS
 fpm build [--profile PROF] [--flag FFLAGS] [--compiler COMPILER_NAME] [-list]

 fpm build --help|--version

DESCRIPTION
 The "fpm build" command
    o Fetches any dependencies
    o Scans your sources
    o Builds them in the proper order

 The Fortran source files are assumed by default to be in
    o src/     for modules and procedure source
    o app/     main program(s) for applications
    o test/    main program(s) and support files for project tests
    o example/ main program(s) for example programs
 Changed or new files found are rebuilt. The results are placed in
 the build/ directory.

 Non-default pathnames and remote dependencies are used if
 specified in the "fpm.toml" file.

OPTIONS
 --profile PROF    selects the compilation profile for the build.
                   Currently available profiles are "release" for
                   high optimization and "debug" for full debug options.
                   If --flag is not specified the "debug" flags are the
                   default.
 --flag  FFLAGS    selects compile arguments for the build. These are
                   added to the profile options if --profile is specified,
                   else these options override the defaults.
                   Note object and .mod directory locations are always
                   built in.
 --compiler   COMPILER_NAME  Specify a compiler name. The default is
                           "gfortran" unless set by the environment
                           variable FPM_COMPILER.
 --list       list candidates instead of building or running them
 --show-model show the model and exit (do not build)
 --help       print this help and exit
 --version    print program version information and exit

EXAMPLES
 Sample commands:

  fpm build                   # build with debug options
  fpm build --profile release # build with high optimization
\end{verbatim}

The simplest and most common use of fpm involves performing a process
like the following:

\begin{enumerate}
\def\labelenumi{\arabic{enumi}.}
\tightlist
\item
  Execute \texttt{fpm\ new} to create a new project.
\item
  Make edits to existing files, and/or create new files in the
  \texttt{app}, \texttt{src}, and/or \texttt{test} directories.
\item
  Execute \texttt{fpm\ test} and/or \texttt{fpm\ run} to ensure that the
  code compiles and executes as expected.
\item
  Edit the \texttt{fpm.toml} file to add new dependencies.
\item
  Publish your project to a repository hosting service such as GitLab,
  GitHub, or any other publicly accessible git server.
\item
  Notify the maintainers of the fpm registry of your new package, to
  allow other users to easily find and use your project.
\end{enumerate}

The primary venue to interact with a Fortran project is the command line
interface of fpm. All common tasks are available via subcommands that
provide reasonable default choices and that can be customized using
flags and/or manifest settings.

\hypertarget{currently-supported-functionality}{%
\subsection{Currently supported
functionality}\label{currently-supported-functionality}}

By default, fpm assumes a specific directory layout for the source files
to be placed in. It automatically finds library sources in the
\texttt{src} directory, applications in the \texttt{app} directory, and
unit tests in the \texttt{test} directory. However, the default behavior
can be overridden by editing the package manifest--a configuration file
in TOML format that resides in the root directory of the project. TOML
is a widely used configuration language that helps minimize the burden
on the programmer to write build files. Fpm also allows creating a new,
skeleton project via the command \texttt{fpm\ new}. One can build, run,
and test a project using the commands \texttt{fpm\ build},
\texttt{fpm\ run}, and \texttt{fpm\ test}, respectively. It is also
possible to install a project via \texttt{fpm\ install}, which performs
the build step if necessary, and makes the project executables available
in user's path.

Internally, fpm builds a detailed model of a Fortran package and its
dependencies by scanning the source files found in default locations and
processing the information provided in a package manifest file and at
the command line. This internal model captures the types of Fortran
sources that are present, their module/submodule interdependencies, and
in which order to invoke the compiler to successfully build the project.
Fpm can thus compile large, complex projects that consist of both
Fortran and C sources, automatically and in parallel where possible. It
contains recommended default sets of compiler flags for \texttt{debug}
and \texttt{release} profiles across several commonly available Fortran
compilers. This is particularly helpful when working with multiple
compilers and different profiles, where fpm automatically separates
build outputs and performs incremental re-compilation when necessary.

Fpm's role as a package manager is perhaps the most innovative aspect,
allowing Fortran developers to effortlessly incorporate third-party
Fortran packages in their own project. It currently supports a
decentralized distribution environment in which project dependencies are
specified as remote \texttt{git} repositories in the project manifest
file. Versioning of project dependencies can be pinned in a
straightforward manner by specifying a commit or tag. Specifying a
dependency in the fpm manifest file requires a single entry, for
example:

\begin{verbatim}
[dependencies]
toml-f = { git = "https://github.com/toml-f/toml-f", tag = "v0.2.1" }
\end{verbatim}

Once added to the manifest, fpm will fetch the remote dependency and
incorporate its library sources into the internal model such that its
modules and procedures can be imported as if they were within the
current package. Likewise, publishing an fpm package as a public git
repository makes it immediately available as a dependency for use in any
other fpm package.

Users can customize various aspects of fpm behavior, for example:

\begin{itemize}
\tightlist
\item
  Specify the compiler and compiler options via command line flags
\item
  Use external module dependencies (such as MPI, NetCDF (Rew and Davis
  1990), and similar) by specifying paths on the command line
\item
  Specify non-default locations for main programs via the manifest
  (\texttt{fpm.toml})
\item
  Specify external libraries, such as BLAS or LAPACK, to be linked via
  the manifest
\end{itemize}

The fpm registry currently lists 32 Fortran packages that can be
automatically downloaded, compiled, and imported in user code using
fpm's dependency feature.

\hypertarget{future-functionality}{%
\subsection{Future functionality}\label{future-functionality}}

While the core functionality of fpm is complete and is gaining traction
with users, many other features are planned for future development:

\begin{itemize}
\tightlist
\item
  Generating configuration files for other build systems like Make,
  CMake, Conda, and others
\item
  Searching for packages in the central fpm package registry
\item
  Better support to specify compiler options
\item
  A plug-in architecture to allow easy development of new capabilities
  as separate packages
\item
  Standardized support for preprocessors
\item
  Better support for external dependencies, such as MPI, NetCDF, and
  other common dependencies
\end{itemize}

Suggestions for others are always welcome and can be submitted at
\url{https://github.com/fortran-lang/fpm/issues}.

\hypertarget{fortran-proposals-incubator-repository}{%
\section{Fortran proposals incubator
repository}\label{fortran-proposals-incubator-repository}}

Fortran-lang efforts stemmed from the
\href{https://github.com/j3-fortran/fortran_proposals}{Fortran proposals
repository} that we created in October 2019 with the aim to allow
anybody to easily discuss new ideas for the Fortran Standards Committee.
Quickly the wide Fortran community started providing feedback and the
repository also served as a way to organize our initial community. After
Fortran-lang was created, the community organized around it, and the
proposals repository is now used specifically to suggest changes to the
language itself.

This repository has been successful and we hope that the community will
use it to provide feedback to the Committee and participate more in the
standardization process. To date, 14 members of the Fortran Standards
Committee have participated in the discussions, and 180 proposals have
been initiated by the community. Many of these proposals motivated the
creation of the Fortran Standard Library. Several others, such as adding
support for modules as namespaces, setting default values for optional
dummy arguments, and improving the API of the Fortran 202X
\texttt{split} intrinsic, have been refined and presented at the
committee meetings. The \texttt{split} intrinsic proposal has been voted
on by the Committee and resulted in the changes to the planned API of
that addition to the Standard. This is direct evidence of the
effectiveness of the Fortran proposals repository.

Another evidence of success is the tighter integration of the community
with the Standards Committee. To date, two new members (Richardson and
Curcic) have joined J3 (PL22.3, the US body of the Standards Committee)
as the direct result of their participation with the proposals
repository and engaging with the Committee. The level of participation
from end-users, the number of proposals submitted, and the new members
that joined the Committee all serve as a record of success of this
GitHub repository.

\hypertarget{lfortran}{%
\section{LFortran}\label{lfortran}}

Another element that has sorely been missing from the Fortran ecosystem
is an interpreter that would allow interactive development for rapid
prototyping. \href{https://lfortran.org}{LFortran} is a modern
open-source (BSD licensed) interactive Fortran compiler built on top of
LLVM. It can execute user's code interactively to allow exploratory work
(like Python, MATLAB, or Julia) as well as compile to binaries to run
user's code on modern architectures such as multi-core CPUs and GPUs.

The initial prototype was written in Python and was successfully shown
to work interactively in a Jupyter notebook. After the public
\href{https://lfortran.org/blog/2019/04/why-we-created-lfortran/}{announcement},
it received positive response from the community.

Since then, we have rewritten it in C++ for speed and robustness. It has
several backends:

\begin{itemize}
\tightlist
\item
  The LLVM backend is used to compile to binaries as well as interactive
  use in Jupyter or a command line REPL
\item
  The C++ backend can translate Fortran code into readable C++
\item
  The x86 backend can generate x86 machine code directly which allows
  fast compilation.
\end{itemize}

As of September 2021, the parser can parse all of Fortran 2018 syntax,
and the compiler can compile simple projects to binaries. These features
also work interactively. We are working toward releasing a Minimum
Viable Product (MVP) that compiles several third-party projects that use
most of Fortran 95 features.

After releasing the MVP, we will implement all Fortran 2018 features,
add more backends, and use LFortran as an infrastructure for other
Fortran tools, such as automatic formatters, documentation generators,
automatic wrappers to other languages (Python, C++, Julia, etc.),
language servers for code editors (VSCode, Vim, Emacs), etc. We also
plan to use the Jupyter notebooks powered by the LFortran kernel to
create interactive tutorials for Fortran, hosted at fortran-lang.org.

\hypertarget{fortran-lang.org}{%
\section{fortran-lang.org}\label{fortran-lang.org}}

\begin{figure}
\centering
\includegraphics{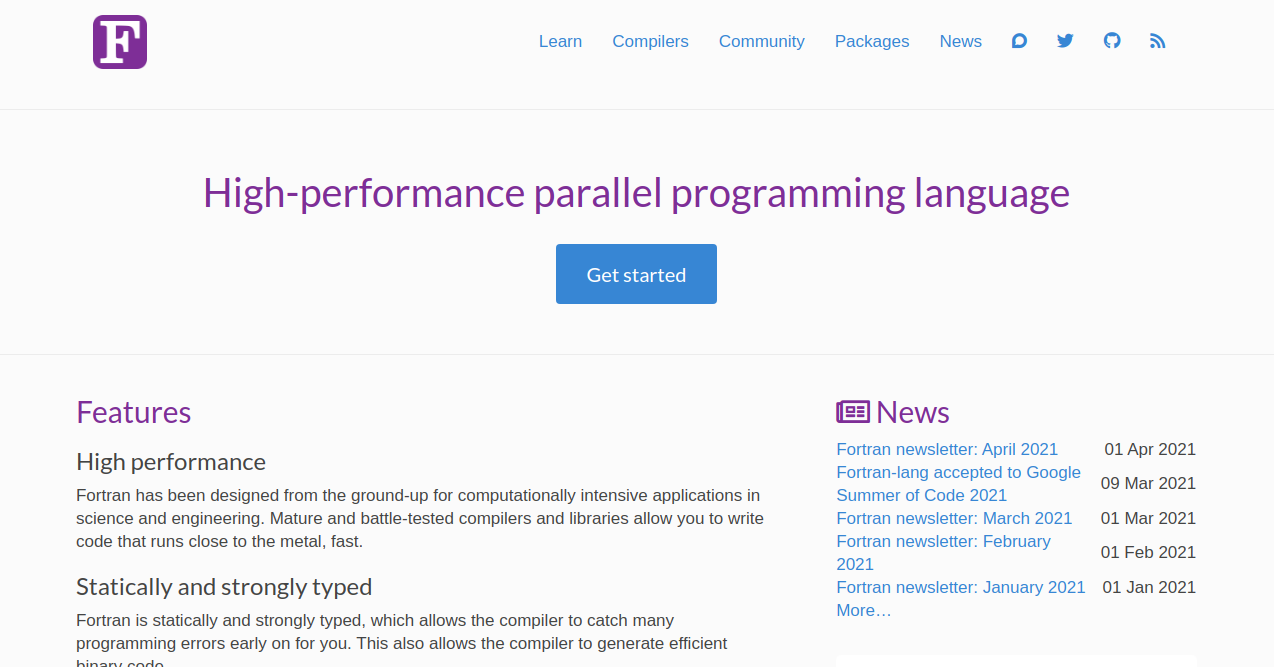}
\caption{Screenshot of the fortran-lang.org website landing
page.\label{fig:fortran-lang-website}}
\end{figure}

In April 2020, another software project was started under the
Fortran-lang umbrella: A community-curated website to serve as the home
of Fortran on the internet (\autoref{fig:fortran-lang-website}).

The website, located at \url{https://fortran-lang.org}, consists of five
components:

\begin{enumerate}
\def\labelenumi{\arabic{enumi}.}
\tightlist
\item
  A Learning page that includes community-curated tutorials
  (mini-books), links to other Fortran resources on the internet, and an
  extensive list of Fortran books and free Fortran courses. The
  community-curated mini-books currently include a ``quick-start''
  Fortran tutorial, a tutorial on building programs, and a tutorial on
  setting up a Fortran development environment. To help improve Fortran
  user acquisition, these tutorials are especially geared toward
  newcomers to the language.
\item
  A Compilers page that lists all Fortran compilers that are currently
  active, commercial and open source alike.
\item
  A Community page that provides links and activity from the
  Fortran-lang projects as well as attributing all GitHub contributors
  to the Fortran-lang project repositories.
\item
  A Package index that provides a community-curated list of, and a
  search bar for, Fortran libraries and applications, tagged by
  category. Anybody can submit a package for inclusion in the index,
  however, there are quality criteria that a package must satisfy, such
  as to have an open source license, to have user documentation, and
  that its core functionality is fully implemented.
\item
  A monthly newsletter in which we report on the progress at the
  beginning of each month.
\end{enumerate}

The website is entirely open source, hosted at
\url{https://github.com/fortran-lang/fortran-lang.org/}, and we welcome
contributions from the community. Most website content is written in the
easy-to-learn Markdown language and the structure is set up such that
adding new content like tutorials is straightforward.

While we do not know the number of visitors to the website due to the
absence of web user trackers, there are at least two signs of early
success of the website. First, the search engine rankings have been
steadily improving during the past year. As of April 2021, the rank of
the result being \url{https://fortran-lang.org} given the search term
``fortran'', while in private-browsing mode, is \#2 for Google, and \#1
for Bing, DuckDuckGo, and SearchEncrypt search engines. Second, the
Fortran page on Wikipedia has been edited by their moderators to include
a URL to the Fortran-lang website. Both the search engine rankings and
the listing on Wikipedia are important because they expand the surface
area over which new users can discover and learn more about Fortran,
thus helping to ensure Fortran's long term well being.

\begin{figure}
\centering
\includegraphics{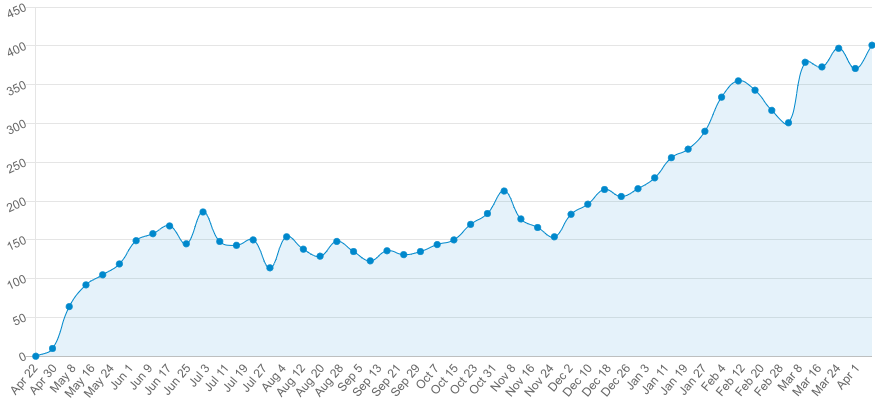}
\caption{Fortran Discourse daily unique visitors.\label{fig:discourse}}
\end{figure}

In addition to the main website, another new and dynamic web presence
for Fortran comes in the form of a modern web forum for discussing all
things Fortran: \url{https://fortran-lang.discourse.group}. The forum is
run on the popular Discourse platform which is used by many programming
language communities thanks to its many invaluable features, including:
easy Markdown formatting of posts, including code blocks; categorization
of threads; user tagging; direct messages; and post moderation. The
Fortran Discourse has quickly grown to nearly 300 registered users
during its first year, and now averages around 4000 page views per day
and 200 new posts per week (Figure \ref{fig:discourse}). Importantly,
the Discourse aims to be an open and welcoming community for all,
beginners and experts alike, through which users may seek help, share
knowledge, and keep up-to-date with Fortran-lang projects and events.
Similarly, a new mailing list
(\href{mailto:fortran-lang@groups.io}{\nolinkurl{fortran-lang@groups.io}})
and Twitter account (@fortranlang) are also provided as a low-bandwidth
alternative for users who wish to be kept informed of Fortran-lang
activity and news more directly.

A successful step toward bringing-together the Fortran community has
been the organization of monthly video calls which are open to all and
advertised via the Discourse, mailing list, and Twitter account. These
meetings allow real-time discussion on the various Fortran-lang projects
presented herein and are particularly beneficial in pushing forward
important decisions with advice and support from the wider community.

\hypertarget{summary}{%
\section{Summary}\label{summary}}

A new open source community called Fortran-lang was formed to address
the lack of modern Fortran tooling such as a comprehensive standard
library, a dedicated build system and package manager, and an
interactive compiler. Since December 2019, it has attracted code
contributions from over 60 participants across several projects.
Fortran-lang's key projects are the Fortran Standard Library (stdlib),
Fortran Package Manager (fpm), and the Fortran website and associated
online platforms. The standard library aims to provide the most common
tools a programmer may need in their daily work, akin to standard
libraries of other programming languages. It currently provides 11
modules with utilities from math and statistics to strings and I/O. Fpm
is both a build system and a package manager, with the main goal to
improve user experience. To date, the key functionalities have been
implemented---creating a new Fortran project, building the library, and
building and running the tests and executables. Fpm also automatically
fetches and builds fpm-enabled dependencies, allowing the programmer to
focus on their library or application development. LFortran is a modern
and interactive open source compiler that will eventually integrate with
other Fortran-lang projects. Finally, the Fortran website provides
community-curated tutorials, compiler resources, and a Fortran package
index. The website, and its associated Discourse forum, mailing list,
and newsletter aim to bring Fortran programmers from around the world
together in a friendly, inclusive, and moderated setting. Together with
the Standard Committees' efforts, we hope that these projects will serve
as a foundation for a stronger, richer, and more accessible Fortran for
the 21st century.

\textbf{Acknowledgements.} We thank all open source contributors who
worked hard across all Fortran-lang projects:
\input{people} and all other participants who engaged
in fruitful discussions on Discourse, Twitter, and GitHub issues. We
thank the Google Summer of Code program that sponsored six summer
students (one through NumFOCUS) that worked on stdlib, fpm, and LFortran
projects. We also thank J3 and WG-5 members who encouraged and supported
our work on Fortran-lang. Ondřej Čertík's work was partially supported
by the U.S. Department of Energy through the Los Alamos National
Laboratory. Los Alamos National Laboratory is operated by Triad National
Security, LLC, for the National Nuclear Security Administration of U.S.
Department of Energy (Contract No.~89233218CNA000001). All other
Fortran-lang projects have been built by 100\% volunteer effort.

\hypertarget{references}{%
\section*{References}\label{references}}
\addcontentsline{toc}{section}{References}

\hypertarget{refs}{}
\begin{CSLReferences}{1}{0}
\leavevmode\vadjust pre{\hypertarget{ref-Aguilar2018}{}}%
Aguilar, Ignacio, Shogo Tsuruta, Yutaka Masuda, Daniela Lourenco, Andrés
Legarra, and Ignacy Misztal. 2018. {``Blupf90 Suite of Programs for
Animal Breeding with Focus on Genomics.''} \emph{Proceedings of the
World Congress on Genetics Applied to Livestock Production} Methods and
Tools - Software: 751.

\leavevmode\vadjust pre{\hypertarget{ref-Apra2020}{}}%
Apra, Edoardo, Eric J Bylaska, Wibe A De Jong, Niranjan Govind, Karol
Kowalski, Tjerk P Straatsma, Marat Valiev, et al. 2020. {``{NWChem:
Past, present, and future}.''} \emph{The Journal of Chemical Physics}
152 (18): 184102.

\leavevmode\vadjust pre{\hypertarget{ref-Aradi2020}{}}%
Aradi, B. 2020. \emph{Fypp: Python-Powered Fortran Metaprogramming}.
GitHub. \url{https://github.com/aradi/fypp}.

\leavevmode\vadjust pre{\hypertarget{ref-Backus_1957}{}}%
Backus, J. W., H. Stern, I. Ziller, R. A. Hughes, R. Nutt, R. J. Beeber,
S. Best, et al. 1957. {``The {FORTRAN} Automatic Coding System.''} In
\emph{Papers Presented at the February 26-28, 1957, Western Joint
Computer Conference: Techniques for Reliability on - {IRE}-{AIEE}-{ACM}
{\textquotesingle}57 (Western)}, 188--98. {ACM} Press.
\url{https://doi.org/10.1145/1455567.1455599}.

\leavevmode\vadjust pre{\hypertarget{ref-Certik2013}{}}%
Čertík, Ondřej, John E. Pask, and Jiří Vackář. 2013. {``Dftatom: A
Robust and General Schrödinger and Dirac Solver for Atomic Structure
Calculations.''} \emph{Computer Physics Communications} 184 (7):
1777--91. \url{https://doi.org/10.1016/j.cpc.2013.02.014}.

\leavevmode\vadjust pre{\hypertarget{ref-Curcic_2019}{}}%
Curcic, M. 2019. {``A Parallel Fortran Framework for Neural Networks and
Deep Learning.''} \emph{{ACM} {SIGPLAN} Fortran Forum} 38 (1): 4--21.
\url{https://doi.org/10.1145/3323057.3323059}.

\leavevmode\vadjust pre{\hypertarget{ref-Diaz_2021}{}}%
Diaz, Javier, Marco Pinna, Andrei V. Zvelindovsky, and Ignacio
Pagonabarraga. 2021. {``Parallel Hybrid Simulations of Block Copolymer
Nanocomposites Using Coarray Fortran.''} \emph{Macromolecular Theory and
Simulations}, March, 2100007.
\url{https://doi.org/10.1002/mats.202100007}.

\leavevmode\vadjust pre{\hypertarget{ref-Donovan2015}{}}%
Donovan, A. A., and B. W. Kernighan. 2015. \emph{The Go Programming
Language}. Addison-Wesley Professional.

\leavevmode\vadjust pre{\hypertarget{ref-Fehr2018}{}}%
Fehr, H., and F. Kindermann. 2018. \emph{Introduction to Computational
Economics Using Fortran}. Oxford University Press.

\leavevmode\vadjust pre{\hypertarget{ref-Garain_2015}{}}%
Garain, Sudip, Dinshaw S. Balsara, and John Reid. 2015. {``Comparing
Coarray Fortran ({CAF}) with {MPI} for Several Structured Mesh {PDE}
Applications.''} \emph{Journal of Computational Physics} 297
(September): 237--53. \url{https://doi.org/10.1016/j.jcp.2015.05.020}.

\leavevmode\vadjust pre{\hypertarget{ref-Gropp_1996}{}}%
Gropp, William, Ewing Lusk, Nathan Doss, and Anthony Skjellum. 1996.
{``A High-Performance, Portable Implementation of the {MPI} Message
Passing Interface Standard.''} \emph{Parallel Computing} 22 (6):
789--828. \url{https://doi.org/10.1016/0167-8191(96)00024-5}.

\leavevmode\vadjust pre{\hypertarget{ref-Harris2020}{}}%
Harris, Charles R, K Jarrod Millman, Stéfan J van der Walt, Ralf
Gommers, Pauli Virtanen, David Cournapeau, Eric Wieser, et al. 2020.
{``{Array programming with NumPy}.''} \emph{Nature} 585 (7825): 357--62.

\leavevmode\vadjust pre{\hypertarget{ref-Harrison1996}{}}%
Harrison, W. J., and K. R. Pearson. 1996. {``Computing Solutions for
Large General Equilibrium Models Using GEMPACK.''} \emph{Computational
Economics} 9: 83--127. \url{https://doi.org/10.1007/BF00123638}.

\leavevmode\vadjust pre{\hypertarget{ref-Fortran2018}{}}%
ISO/IEC JTC 1/SC 22/WG 5. 2018. {``Information Technology ---
Programming Languages --- Fortran --- Part 1: Base Language.''}
Technical Report. International Organization for Standardization.

\leavevmode\vadjust pre{\hypertarget{ref-JonesCrabtree1988}{}}%
Jones, Russell K., and Tracy Crabtree. 1988. \emph{Fortran Tools for
{VAX/VMS} and {MS-DOS}}. John Wiley \& Sons.

\leavevmode\vadjust pre{\hypertarget{ref-klabnik2019rust}{}}%
Klabnik, Steve, and Carol Nichols. 2019. \emph{The Rust Programming
Language (Covers Rust 2018)}. No Starch Press.

\leavevmode\vadjust pre{\hypertarget{ref-MacMackin2018}{}}%
MacMackin, Christopher. 2018. \emph{FORD} (version v6.0.0). Zenodo.
\url{https://doi.org/10.5281/zenodo.1422473}.

\leavevmode\vadjust pre{\hypertarget{ref-Mozdzynski_2015}{}}%
Mozdzynski, George, Mats Hamrud, and Nils Wedi. 2015. {``A Partitioned
Global Address Space Implementation of the European Centre for Medium
Range Weather Forecasts Integrated Forecasting System.''} \emph{The
International Journal of High Performance Computing Applications} 29
(3): 261--73. \url{https://doi.org/10.1177/1094342015576773}.

\leavevmode\vadjust pre{\hypertarget{ref-Plauger1992}{}}%
Plauger, P. J. 1992. \emph{{The Standard C Library}}. Prentice Hall.

\leavevmode\vadjust pre{\hypertarget{ref-Powers2017}{}}%
Powers, Jordan G., Joseph B. Klemp, William C. Skamarock, Christopher A.
Davis, Jimy Dudhia, David O. Gill, Janice L. Coen, et al. 2017. {``{The
Weather Research and Forecasting Model: Overview, System Efforts, and
Future Directions}.''} \emph{Bulletin of the American Meteorological
Society} 98 (8): 1717--37.
\url{https://doi.org/10.1175/BAMS-D-15-00308.1}.

\leavevmode\vadjust pre{\hypertarget{ref-Reid_2020}{}}%
Reid, John, Bill Long, and Jon Steidel. 2020. {``History of Coarrays and
{SPMD} Parallelism in Fortran.''} \emph{Proceedings of the {ACM} on
Programming Languages} 4 ({HOPL}): 1--30.
\url{https://doi.org/10.1145/3386322}.

\leavevmode\vadjust pre{\hypertarget{ref-Rew1990}{}}%
Rew, R., and G. Davis. 1990. {``NetCDF: An Interface for Scientific Data
Access.''} \emph{IEEE Computer Graphics and Applications} 10 (4):
76--82. \url{https://doi.org/10.1109/38.56302}.

\leavevmode\vadjust pre{\hypertarget{ref-sanner1999python}{}}%
Sanner, Michel F et al. 1999. {``Python: A Programming Language for
Software Integration and Development.''} \emph{J Mol Graph Model} 17
(1): 57--61.

\leavevmode\vadjust pre{\hypertarget{ref-Schonfelder1996}{}}%
Schonfelder, J. L. 1996. {``Technical Deficiencies with Fortran 90.''}
\emph{Computer Standards \& Interfaces} 18 (4): 295--302.
\url{https://doi.org/10.1016/0920-5489(96)01005-7}.

\leavevmode\vadjust pre{\hypertarget{ref-Schonfelder1991}{}}%
Schonfelder, Lawrie. 1991. {``The Future Evolution of Fortran.''} In
\emph{ACM SIGPLAN Fortran Forum}, 10:16--17. 2. ACM New York, NY, USA.

\leavevmode\vadjust pre{\hypertarget{ref-sharma2017}{}}%
Sharma, Anuj, and Irene Moulitsas. 2017. {``MPI to Coarray Fortran:
Experiences with a CFD Solver for Unstructured Meshes.''}
\emph{Scientific Programming}.
\url{https://doi.org/10.1155/2017/3409647}.

\leavevmode\vadjust pre{\hypertarget{ref-Shterenlikht2018}{}}%
Shterenlikht, Anton, Lee Margetts, and Luis Cebamanos. 2018.
{``{Modelling fracture in heterogeneous materials on HPC systems using a
hybrid MPI/Fortran coarray multi-scale CAFE framework}.''}
\emph{Advances in Engineering Software} 125: 155--66.

\leavevmode\vadjust pre{\hypertarget{ref-Skamarock2021}{}}%
Skamarock, William C., Hing Ong, and Joseph B. Klemp. 2021. {``{A Fully
Compressible Nonhydrostatic Deep-Atmosphere Equations Solver for
MPAS}.''} \emph{Monthly Weather Review} 149 (2): 571--83.
\url{https://doi.org/10.1175/MWR-D-20-0286.1}.

\leavevmode\vadjust pre{\hypertarget{ref-stroustrup2013c}{}}%
Stroustrup, Bjarne. 2013. \emph{{The C++ programming language}}. Pearson
Education.

\leavevmode\vadjust pre{\hypertarget{ref-Vandenplas2020}{}}%
Vandenplas, Jeremie, Herwin Eding, Maarten Bosmans, and Mario P. L.
Calus. 2020. {``Computational Strategies for the Preconditioned
Conjugate Gradient Method Applied to {ssSNPBLUP}, with an Application to
a Multivariate Maternal Model.''} \emph{Genetics Selection Evolution} 52
(1): 24. \url{https://doi.org/10.1186/s12711-020-00543-9}.

\leavevmode\vadjust pre{\hypertarget{ref-Virtanen2020}{}}%
Virtanen, Pauli, Ralf Gommers, Travis E Oliphant, Matt Haberland, Tyler
Reddy, David Cournapeau, Evgeni Burovski, et al. 2020. {``{SciPy 1.0:
fundamental algorithms for scientific computing in Python}.''}
\emph{Nature Methods} 17 (3): 261--72.

\end{CSLReferences}

\end{document}

%% file: authors.tex
\author[a]{Milan Curcic}
\author[b]{Ondřej Čertík}
\author[c]{Brad Richardson}
\author[d]{Sebastian Ehlert}
\author[e]{Laurence Kedward}
\author[f]{Arjen Markus}
\author[g]{Ivan Pribec}
\author[h]{Jérémie Vandenplas}

\affil[a]{University of Miami}
\affil[b]{Los Alamos National Laboratory}
\affil[c]{Archaeologic Inc.}
\affil[d]{University of Bonn}
\affil[e]{University of Bristol}
\affil[f]{Deltares}
\affil[g]{Technical University of Munich}
\affil[h]{Wageningen University and Research}

%% file: people.tex
Aaron Graham,
Aaron Meurer,
acferrad,
Adam Denchfield,
Aditya Shukla,
Adriano Amaricci,
Agustin Forero,
Alabi Olasunkanmi Matthew,
Alexis Perry-Holby,
Alfio Lazzaro,
Alin Marin Elena,
Aman Godara,
Amir Shahmoradi,
Amit Kumar,
Anar Yusifov,
Andre Mikulec,
Andre Smit,
andrew\_4619,
Andy Malcolm,
Aniruddha Das,
Ankit Pandey,
Antoine Levitt,
Arijit Banerjee,
Armin Wehrfritz,
Arttu Räsänen,
arunningcroc,
Asdrubal Lozada-Blanco,
ashe,
Ashish Panigrahi,
Ashok Jammi,
Ashutosh Pandey,
Ashwin Vishnu,
Atias,
Avinash Vijayvargiya,
Bálint Aradi,
Beliavsky,
Bharat Mahajan,
Bhuvnesh Vashishat,
Bill Long,
Bishal Das,
bitbag,
Bob Apthorpe,
Bojan Niceno,
Brian Callahan,
Brian Cornille,
Bryce Adelstein Lelbach,
bwe,
Carl Burkert,
Carlos Une,
catulus,
CConde,
Chetan Karwa,
Chris Elrod,
Chris MacMackin,
Clive Page,
Conor Hoekstra,
cradesto,
CYehLu,
Cyrille Lavigne,
Damian Rouson,
dannytoearth,
Dauptain,
dav05,
David Bentlage,
David Duffy,
David Lauvergnat,
David McParland,
David Meyer,
David Navia,
David Spurr,
Deepak Kumar,
Dhiganth Rao,
difference-scheme,
Dmitrii Pasechnik,
Dobrodzieju,
Dominic Poerio,
Dominik Gronkiewicz,
D. SundarChand,
E. D. Bakker,
Edger Sterjo,
Elias Rabel,
Emanuele Pagone,
Emilio,
Eugene Epshteyn,
Euler-37,
Evan Voyles,
Ev. Drikos,
Evert Provoost,
Fábio Malacco Moreira,
Fabio Souto de Azevedo,
Fernando Oleo Blanco,
Fernando Pérez,
Filippo M.,
Fran,
Gabriel Brown,
Gagandeep Singh,
Gareth Davies,
Gary Klimowicz,
Gary Styger,
Gilles Gouaillardet,
Greg Hamblin,
haidangwy,
Hank,
Han Tang,
haomiao,
haraldkl,
Harper,
Harris Snyder,
Harsh Bardhan Mishra,
Herve Wozniak,
Himanshu Pandey,
Hongyi Zhao,
hpigula,
Hugo Nava Ramos,
H. Vetinari,
Ian Allison,
Ian Bush,
Ian Chivers,
Ian Giestas Pauli,
Ignacio López de Arbina,
implicitall,
Irvise,
Ivan Morozov,
Izaak "Zaak" Beekman,
Jacob Marks,
Jacob Williams,
Jakub Jelínek,
Jamie,
Jane Sleightholme,
Jason Blevins,
Jason H. Nicholson,
Jason Miller,
jbdv,
Jeff Armstrong,
Jeff Hammond,
Jellby,
Jerry DeLisle,
Jim Fisher,
Joe Schoonover,
Johannes Gebert,
Johannes\_R,
Johann Weber,
John Alexiou,
John Collins,
John Siegrist,
John Urban,
jomyer,
Jordan Sissel,
José Mª Escartín,
Joseph Schoonover,
Juan,
Juan Fiol,
Julian Henneberg,
Jürgen Reuter,
jurnix,
Justin Dedewanou,
Kai Gao,
Karan Kapoor,
Karl (rcs),
Katherine Holcomb,
Ketan Patil,
Kiran Chandramohan,
Kjell Jorner,
Laetis,
Laurent Piau,
Leandro Martinez,
Lenore M Mullin,
Leonard Reuter,
Leon Foks,
Liam Portanier Mifsud,
lm\_lxt,
Lori A. Burns,
Luca,
Luca Argenti,
ludovico,
Luo Heng,
Lysandros,
Mark Johnson,
Mark LeAir,
Mark Wieczorek,
Marshall Ward,
Martin Diehl,
Martín D. Maas,
Martin Jambor,
Mason,
Matt Alex,
Matthew Mitchell,
Matthias Bussonnier,
Matt Houston,
Melissa Mendonça,
Michael Hirsch,
Michael Konečný,
Michael Metcalf,
Michael Rudko,
Michael Siehl,
Michael Zingale,
Miguel Raz,
Miguel Zuniga,
mikolajpp,
Milad Sadeghi,
Mitja,
mobius-eng,
Mohd Furquan,
msz59,
Nathaniel Shaffer,
Nathan Sircombe,
nbehrnd,
Neels van Niekerk,
Neil Carlson,
Nick Papior,
Nick Wogan,
Nikola Vitas,
Nocaster60,
Oleg Peil,
Oliver Fuhrer,
Oscar Garcia Cabrejo,
Oscar Smith,
oyvind,
Pablo Santamaria,
paul,
Pedro Costa,
Peter Brady,
Peter Klausler,
Peter Occil,
Peter S. Simon,
Philippe Blain,
Philipp Engel,
Pierre de Buyl,
Pompito,
PrestonMag,
qolin1,
Rachitt Shah,
Rashid Rafeek,
rb,
rcs,
Reinh Bader,
Reuben Budiardja,
Richard Bleikamp,
Richard Weed,
Robert A Magliola,
Robert Cohn,
Robert D Ryne,
Robert Farmer,
Roberto Putzu,
Robert Ryley,
Robert Van Amerongen,
Rohit Goswami,
Ron,
Ron Caplan,
Rong,
Ron Shepard,
Royi,
Salvatore Filippone,
Sam Harrison,
Sam Partee,
Samuel,
Samuel (GF),
Sascha Klawohn,
Scot Halverson,
Sebastiaan Janssens,
Sebastian Costamagna,
Sebastian Müller,
Seif Shebl,
septcolor,
Shahid Maqbool,
ShamanTcler,
Sig Selene,
Silvio Traversaro,
Simon Braß,
Simon Geard,
Srinath Vadlamani,
Stavros Meskos,
Stefan Karpinski,
Stephen Poole,
Steve Kargl,
Steve Lionel,
Steve Westenbroek,
St Maxwell,
Tanweer Ali,
Theodore-Thomas Tsikas,
thierry,
Thirumalai Shaktivel,
Thomas Henlich,
Tim,
Tim Coupmodel,
Tim Keith,
Tim W,
Tiziano Müller,
Tom Clune,
Tomohiro Degawa,
Truong Dang,
Tyler Chang,
Uncle\_Paul,
Utkarsh Shah,
v91\_bleach,
valtterib,
Van Snyder,
victorsndvg,
Vikas Sharma,
Vincent Magnin,
Vipul Parekh,
Vladimír Fuka,
Walt,
Warrick Ball,
waynelapierre,
Wileam Y. Phan,
William Clodius,
Williams Lima,
witenion\_v2,
Wolf Vollprecht,
xushanthu,
Yehan Wasura,
Yuichiro Sakamoto,
Yutaka Masuda,
Zachary Moon,
Zach Jibben,
Zaikun,
zmiimz,
Zuo Zhihua,